# PREPARATION OF $Bi_4Ti_3O_{12}$ THIN FILMS BY A PULSED LASER ABLATION DEPOSITION (PLAD) METHOD AND THEIR APPLICATIONS FOR ULTRAVIOLE DETECTORS

**_A.Fuad[1], Darsikin[2], Irzaman[3], T.Saragi[4], D.Rusdiana[5], H.Saragih[6], A.Kartono[3], M. Barmawi[*], P. Arifin[*] and D. Kurnia[*]_**

[1]*Department of Physics, Malang University*
[2]*Department of Physics, Tadulako University*
[3]*Department of Physics, Bogor Institute of Agriculture*
[4] *Department of Physics, Sam Ratulangi University*
[5]*Department of Physics, Indonesia Education University*
[6]*Department of Physics, Pattimura University*
**Laboratory of Electronic Material Physics, Department of Physics ITB*

**Abstract**

*$Bi_4Ti_3O_{12}$ (BIT) has been considered as material for UV detector since it has optical band gap of about 3.5 eV. Unlike the most used material for this purpose, such as GaN and AlN, BIT is a ferroelectric oxide material. It has shown good detection properties in the UV regions.*

*We have developed a Pulsed Laser Ablation Deposition (PLAD) method for preparation of BIT thin films. The BIT films were grown on $Si/SiO_2/Pt(111)$ substrates and these films exhibit good crystalline properties with preferential orientation in c-axis. A structure of $Al/Bi_4Ti_3O_{12}/Pt(111)/Si(100)$ was fabricated for measurement of spectral distribution of the voltage responsivity. The maximum voltage responsivity and the spectral distribution significantly depend on the films preparations and treatments.*

**Keywords : $Bi_4Ti_3O_{12}$, PLAD and UV detectors.**

## 1. Introduction

Recently, the interest in developing ultraviolet detectors and optoelectronic devices has continuously increased. The most used materials for UV detection are the large gap semiconductors such as GaN or AlN [1,2]. Unfortunately, detectors made from these materials or combinations of them are obtained mostly by expensive method like metalorganic chemical vapor deposition (MOCVD) or molecular beam epitaxy (MBE) [3].

In addition, to obtain good quality thin films for device fabrication is also difficult matter due to the lattice mismatch between the III-V nitrides and the common substrates used for the deposition (e.q. MgO, sapphire or SiC.

The photoelectric properties of Ferroelectric oxide thin films and heterostructures has shown that these materials have good detection properties in the UV region [4,5].

In this paper, we report the development pulsed laser ablation deposition (PLAD) method for fabrication of BIT ferroelectric oxide thin films. The UV detection properties of $Bi_4Ti_3O_{12}$ (BiT) thin films deposited on $Si/SiO_2/Pt$ substrates are also presented.

## 2. Experiment

The BIT films with 200 nm thickness were deposited on $Si/SiO_2/Pt$ substrates. Details of the deposition technique are shown in table 1.

Semitransparent aluminum metal electrodes with area of 1 x 1 $mm^2$ were evaporated on the ferroelectric film surface. The ferroelectric capacitors defined by the top aluminum electrode and the bottom platinum electrode were used to perform photovoltaic measurement in the spectral wavelength region of 150–400 nm.

Table 1. The deposition condition of thin films BIT by the PLAD method.

| Parameter | Condition |
|---|---|
| Substrates | $Si/SiO_2/Pt$ |
| Target | $Bi_4T_3O_{12}$, |
| Deposition temperature | 600 to 700°C |
| Laser | NdYAG |
| Laser power | 225 mJ |
| Wavelength | 355 nm |
| Deposition pressure | 200 mTorr |
| Oxygen flow rate | 10 Sccm |
| Annealing temperature | 650°C |
| Atmosphere annealing | $O_2$ |
| Thickness | 200 nm |

The signal was measured in modulated light by measuring the ac voltage signal across a 100 KΩ. The Schematic diagram of the photovoltaic signal measurement is shown in figure 2. For the photovoltaic measurement, no voltage was applied to the sample and the incident light was modulated at 1000 Hz with a mechanical chopper. The signal was measured with a Stanford Research lock-in amplifier. The light source used for the measurements was UV Xenon lamp model LPS255HR and the spectral distribution measurements were performed using a Spex 1681 spectrometer.

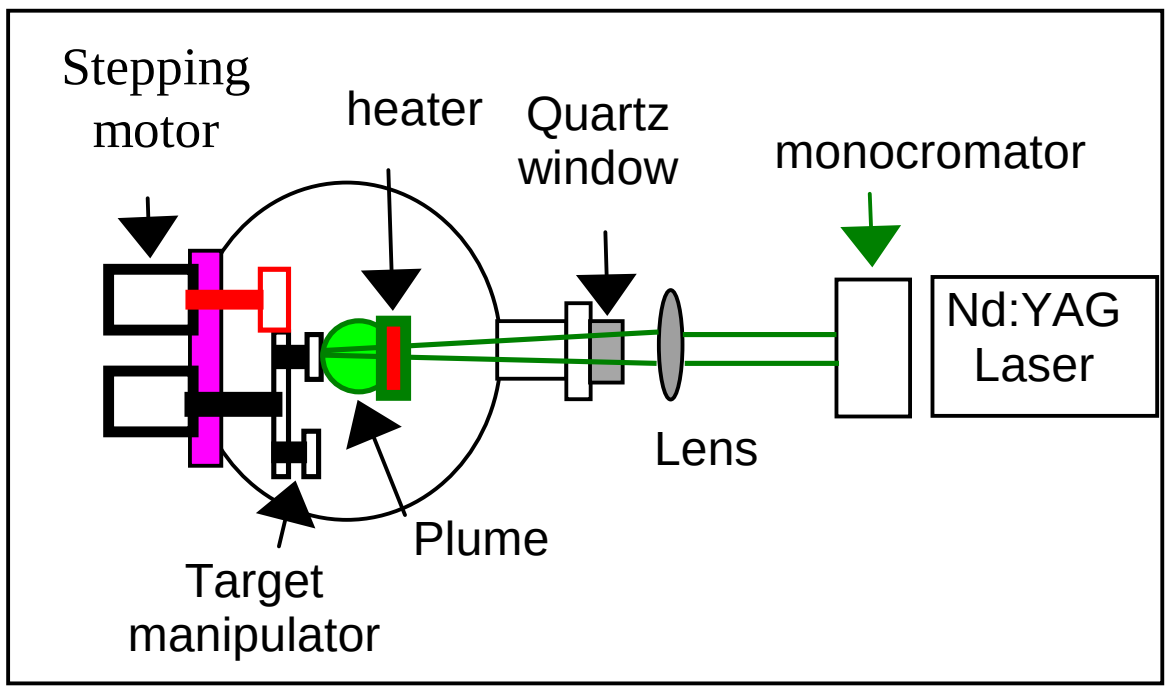

Figure 1. The Schematic diagram of thePLAD system [6]

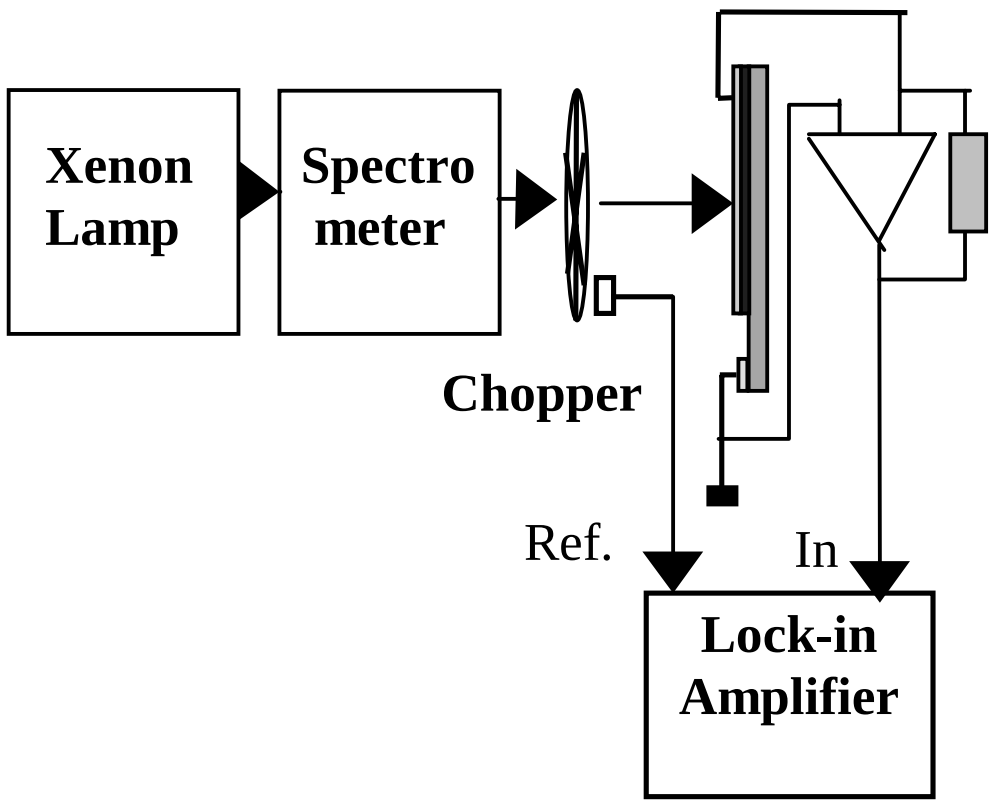

Figure 2. The schematic diagram of the voltage responsivity measurement.

Voltage responsivity was computed using the following formula[7]:

$$R_V = \frac{V}{P_{inc}} \qquad (1)$$

where "V" is the voltage measured cross the load resistance, and $P_{inc}$ is the power of the incident light.

## 3. Results and discussion

Figure 3 shows the spectral distribution of the voltage responsivity measured under modulated light using photovoltaic mode for film annealed at different temperatures. The cut-off wavelength is around 240 nm, but a small tail can be observed up to 390 nm. This tail is probably due to band tails which are present in forbidden gap and act as trapping centers for photogenerated carriers. The band tails can appear due to the translation disorders which are always present in polycrystalline thin films containing oxygen.

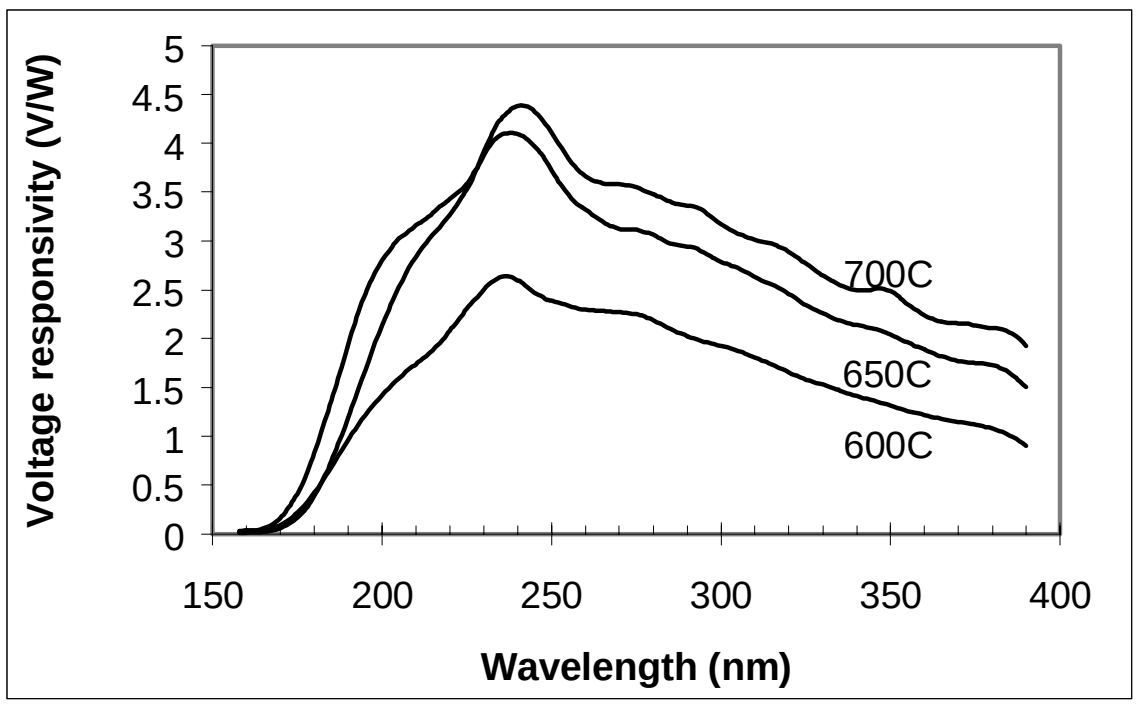

Figure 3. The spectral distribution of the voltage responsivity for thin films BIT annealed at different temperature

The maximum voltage responsivity occurs at wavelength as around 230 nm, and this slightly depends on the annealing temperature of the ferroelectric film. Increasing the deposition temperature from 600 to 700$^{o}$C, the maximum voltage responsivity shift from 230 to 245 nm (Fig.3).

It was also observed that the photovoltaic signal increases with an increase of the deposition temperature of BIT films. By increasing the deposition temperature, the crystalline structure of the film would be improving the ferroelectric properties, i.e., increasing the value of their remanent ferroelectric polarization.

In the real detection system, in order to avoid the errors due to the parasitic light sources, the incident light is usually modulated. Thus, we measured the ac voltage developed across the load resistance $\boldsymbol{R_L}$ = 100 KΩ connected in parallel with the film. The measured signal could originate from a combination of photovoltaic and pyroelectric effect, but we are believe that the photovoltaic effect is dominant as long as the spectral distribution of the voltage responsivity $\boldsymbol{R_V}$ are wavelength dependent (Fig. 3). If the phyroelectric effect is dominant, the spectral distribution should be almost flat.

Figure 4 shows the dependence of the spectral distribution on the deposition temperature of the ferroelectric films for a wavelength of 150–400 nm. Considering that the voltage V measured across the load resistance $\boldsymbol{R_L}$ is given by the product between this resistance and the short-circuit current generated by the ferroelectric film due to the photovoltaic effect.

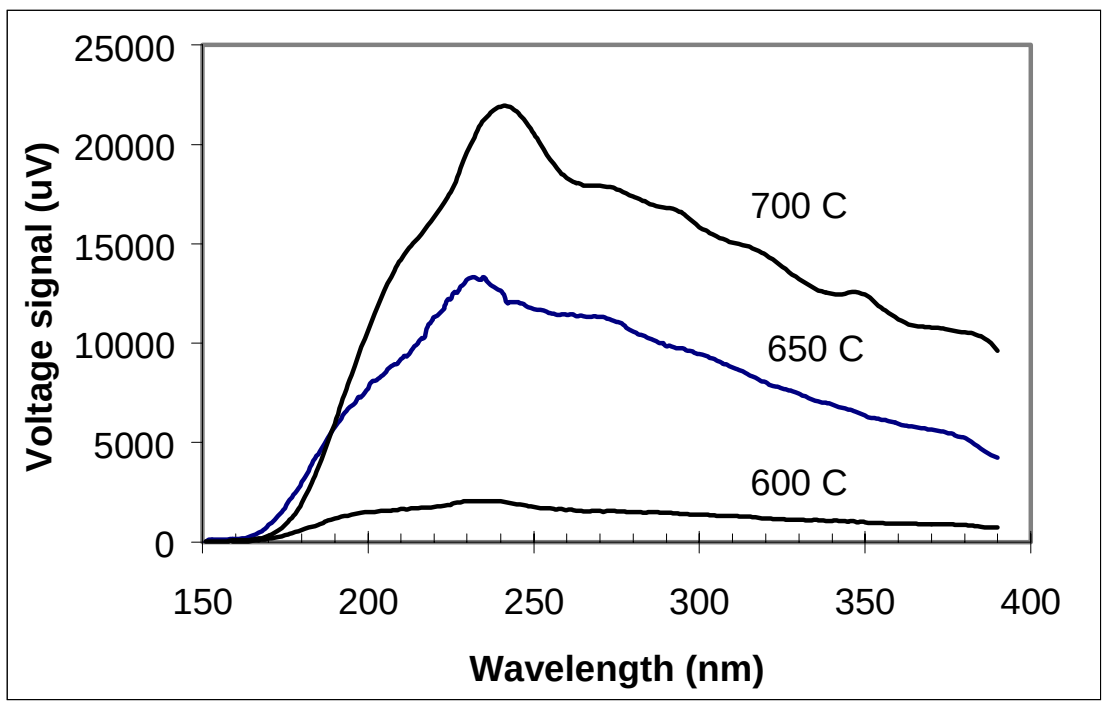

Figure 4. The spectral distribution of the voltage signal across the load resistance $R_L$.

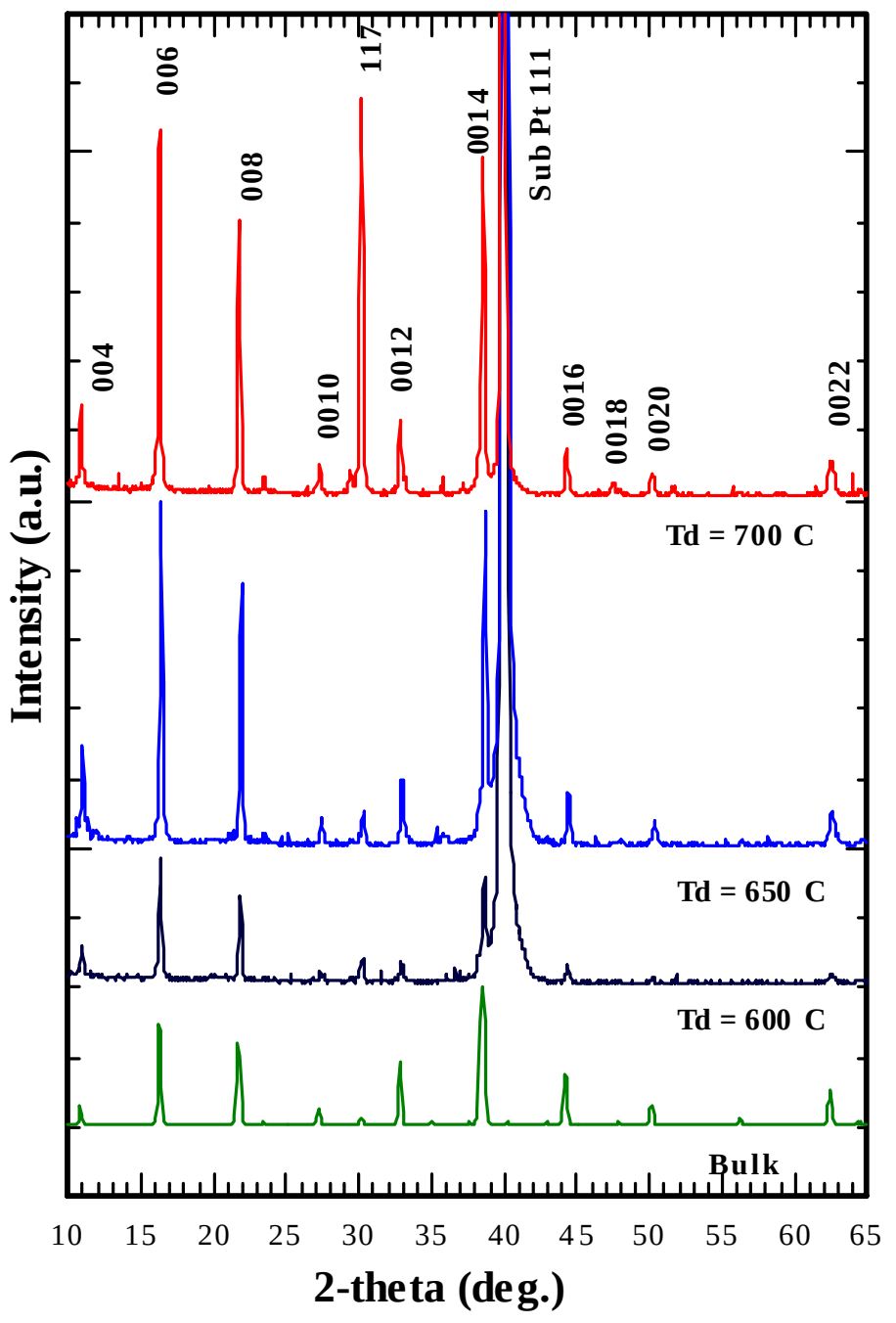

Figure 5. The X-ray diffraction patterns of thin films and bulk BIT grown at different deposition temperature

Figure 5 shows the x-ray difractogram of the target and thin films BIT. These films and bulk exhibit good crystalline properties with preferential orientation in c-axis. These preferential orientations are respected with the maximum voltage responsivity. The voltage responsivity of the BIT thin films increases with the peak intensity of (006) orientation as well as the substrate temperatures.

This voltage responsivity is comparable with the FWHM (006) peak orientation. The FWHM value of the (006) peak orientation were 0.2, 0.18 and 0.16 which was grown at the substrate temperatures 600$^o$C, 650$^o$C and 700$^o$C, respectively.

## 4. Conclusion

The BIT Ferroelectric thin films are good candidates for UV detectors, because high voltage responsivity at the wavelength range from 150 to 400 nm. Thin films BIT grown by PLAD method shows preferential orientation highly *c*-axis. The configuration Si(100)/$SiO_2$/Pt(111)/BIT/Al is very convenient for UV detectors.

## ACKNOWLEDGMENTS

The present work is supported by Laboratory for Electronic Material Physics ITB Bandung.